\documentclass[twocolumn]{IEEEtran}

\usepackage{cite}
\usepackage{url}
\usepackage{graphicx}
\usepackage{amsmath}
\usepackage{amssymb}
\usepackage{wrapfig}
\usepackage{tikz}
\usetikzlibrary{shapes,arrows}
\usetikzlibrary{backgrounds}
\usepackage{cleveref}
\usepackage{float}
\usepackage{sidecap}
\usepackage{placeins}
\usepackage{caption}
\usepackage{tabularx}
\usepackage[shortlabels]{enumitem}
\usepackage{array,multirow,makecell}
\newcolumntype{C}[1]{>{\centering\arraybackslash }b{#1}}

\DeclareMathOperator{\e}{e}
\newcommand{\subscript}[2]{$#1 _ #2$}
\newcommand{\var}{\mathbb{V}}
\newcommand{\mxp}{\mathbb{E}}
\newcommand{\cov}{\textrm{Cov}}
\newcommand\Item[1][]{%
  \ifx\relax#1\relax  \item \else \item[#1] \fi
  \abovedisplayskip=0pt\abovedisplayshortskip=0pt~\vspace*{-\baselineskip}}

\usepackage{mathrsfs}
\usepackage{xcolor}

\begin{document}

\title{The Statistics of the Cross-Spectrum and the Spectrum Average: Generalization to Multiple Instruments}

\author{Antoine Baudiquez, \'{E}ric Lantz, Enrico Rubiola, Fran\c{c}ois Vernotte
\thanks{A. Baudiquez and E. Rubiola are with FEMTO-ST, Department of Time and Frequency, UMR 6174, Universit\'{e} Bourgogne Franche-Comt\'{e}, France. Antoine's ORCID is 0000-0002-7007-5273}
\thanks{E. Lantz is with FEMTO-ST, D\'{e}partement d'Optique P.M. Duffieux, UMR 6174 CNRS, Universit\'{e} Bourgogne Franche-Comt\'{e}, France.}
\thanks{E. Rubiola is also with the Division of Quantum Metrology and Nanotechnology, Istituto Nazionale di Ricerca Metrologica (INRiM), 10135 Turin, Italy. Enrico’s ORCID is 0000-0002-5364-1835}
\thanks{F. Vernotte is with FEMTO-ST, Department  of  Time  and  Frequency, Observatory  THETA, UMR 6174 CNRS, Universit\'{e} Bourgogne Franche-Comt\'{e}, France. Fran\c{c}ois's ORCID is 0000-0002-1645-5873}}

\maketitle

\IEEEpeerreviewmaketitle

\begin{abstract}
This article addresses the measurement of the power spectrum of red noise processes at the lowest frequencies, where the minimum acquisition time is so long that it is impossible to average on a sequence of data record.  Therefore, averaging is possible only on simultaneous observation of multiple instruments.  This is the case of radio astronomy, which we take as the paradigm, but examples may be found in other fields such as climatology and geodesy.\\
We compare the Bayesian confidence interval of the red-noise parameter using two estimators, the spectrum average and the cross-spectrum.  While the spectrum average is widely used, the cross-spectrum using multiple instruments is rather uncommon.  With two instruments, the cross-spectrum estimator leads to the Variance-Gamma distribution. A generalization to $q$ devices is provided, with the example of the observation of millisecond pulsars with 5 radio telescopes.
\end{abstract}


\IEEEpeerreviewmaketitle

\section{Introduction\label{sec1}}
\IEEEPARstart{T}{}he term red noise refers to a variety of processes sharing the property that the power spectral density (PSD) grows at low frequency as $1/f^2$ (Brownian noise) or $1/f^\alpha$, with $\alpha > 2$. We are interested in the estimation of the PSD of such random signals out of the background noise of the instrument in the specific case of very slow phenomena, which take too long acquisition time for the average on a sequence of data sets to be viable. Therefore, averaging out the background is possible only by exploiting simultaneous measurements of the same signal taken with multiple instruments, under the obvious hypothesis that they are independent. The frequency stability of the millisecond pulsars is the example we have in mind. Such rapidly rotating neutron stars, emitting highly stable periodic pulses out of the magnetic poles, rival the best atomic clocks \cite{rawley1987,taylor1991,hartnett2011,manchester2017}. Among other fields, slow phenomena are found in climatology \cite{hansen2012} and geodesy, the latter nowadays measured with Very Large Baseline Interferometry \cite{nothnagel2021}.\\
With the purposes stated in mind, we compare the efficiency of the spectrum average (s.a) and with the cross-spectrum (c-s) measuring the signal with $q$ instruments simultaneously. The s.a estimator is the average of the $q$ observed spectra $S_i$, weighted with the background noise $\sigma_{N,i}^2$ of the $i$-th instrument. The c-s method is the average of the all combinatorial choices of the cross-spectrum $S_{j,i}$, $i \neq j$. The s.a is the classical estimator used in these cases \cite{tiuri1966}, while the c-s is rather uncommon. Data are analyzed with the Bayesian statistics, also known as the inverse problem, which consists of estimating the most probable value of the signal (the slowest spectral components) from the experimental outcomes and their statistical properties. We take the 95\% upper limit as the efficiency criterion. Accordingly, the most efficient estimator is the one that provides the most stringent upper limit with the same data set.\\
Our previous article \cite{baudiquez2020} shows that the Variance-Gamma ($V\Gamma$) distribution is the exact solution for the probability density function (PDF) of the cross-spectrum in the case of two instruments. We generalize the result to the case of the cross spectrum of $q$ instruments, each with its own background noise $\sigma_{N_i}^2$, assessing the confidence interval on the signal level $\sigma_R^2$. Of course, the PDF is no longer a $V\Gamma$, and can only be calculated numerically. The case of equally noisy instruments is simpler, and at first sight similar to that of $q = 2$, but it has no analytical solution.\\
We run a simulation with up to five instruments, inspired to the LEAP experiment \cite{bassa2015}. Such experiment gathers the five largest European radio telescopes (RTs) in order to increase the sensitivity of high-precision pulsar timing. Interestingly, Pulsar Timing Arrays seem a promising option to explore the low-frequency gravity waves crossing our Galaxy \cite{verbiest2021,goncharov2021}.\\
The simulation shows that the s.a is by a small amount more efficient than the c-s, chiefly when the background exceeds the signal. Indeed, this depends on the numerical values. In the end, the use of both estimators may be a wise choice.

\begin{figure}
    \includegraphics[width=\linewidth]{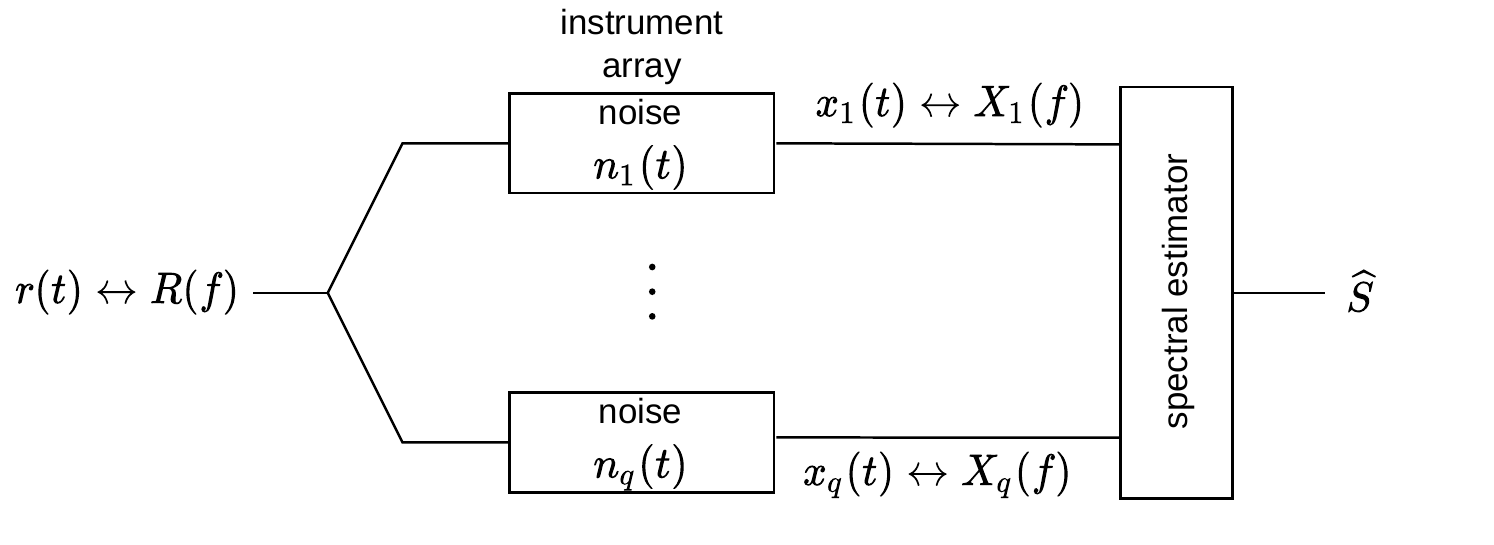}
    \caption{Array of $q$ instruments measuring the signal $r(t)$. Each RT adds a white noise to the output $x(t)$ whose Fourier transform is $X(f)$. Then the estimate $\widehat{S}$ is computed.\label{fig1}}
\end{figure}

\section{Statement of the Problem\label{sec2}}
\subsection{Spectral Measurement\label{sec2.1}}

Let us consider a \textbf{red} noise $r(t)$ which is measured by $q$ independent instruments as shown in Fig. \ref{fig1}. We assume that each instrument adds a white noise $n_i(t)$ to the measurement and that all these white noises are uncorrelated. The output of each channel is then

\begin{equation}
x_i = r + n_i \leftrightarrow X_i = R + N_i
\label{eq:time-freq}
\end{equation}
where the subscript $i$ corresponds to the $i$-th instrument, $\leftrightarrow$ stands for the Fourier transform and inverse Fourier transform pair, lower case is time domain, upper case is frequency domain, and the variables $t$ and $f$ are implied. Let us remind that the Fourier transform of a white noise is a white noise, at least for sampled signals. Indeed even if continuous pure white noise have an infinite power, the Fourier transform for discrete simulation can be defined. A realistic white noise corresponds to a Markov process of the first order, more details about colored noise are given in \cite{kasdin1995}.\\
On the other hand, a red noise can be described as a filtered white noise. Its spectrum is then the product of a white spectrum by a deterministic function; so the random part of a red noise is uncorrelated for each frequency bin. Consequently in term of random variable, working in the frequency domain gives a precious advantage because the Fourier components (frequency bins) are statistically independent unlike the time data.\\
In the following we focus solely on \textbf{one frequency bin}, thanks to energy equipartition it follows,

{\setlength{\arraycolsep}{2pt}
\begin{equation}
    \begin{array}{llll}
        \mathbb{V}\left[N_i\right] &= 2 \mathbb{V} \left[\Re\left[N_i\right]\right] &= 2 \mathbb{V} \left[\Im\left[N_i\right]\right] &= \sigma_{N,i}^2\\[0.3cm]
        \mathbb{V}\left[R\right] &= 2 \mathbb{V} \left[\Re\left[R\right]\right] &= 2 \mathbb{V} \left[\Im\left[R\right]\right] &= \sigma_{R}^2
        \label{eq:variance_factor}
    \end{array}
\end{equation}
where $\mathbb{V}[\cdot]$, $\Re[\cdot]$, $\Im[\cdot]$ respectively denote the variance, the real and imaginary part of the quantity within the brackets.

\subsection{Periodogram and Power Spectral Density\label{sec2.2}}

\begin{figure}
    \includegraphics[width=\linewidth]{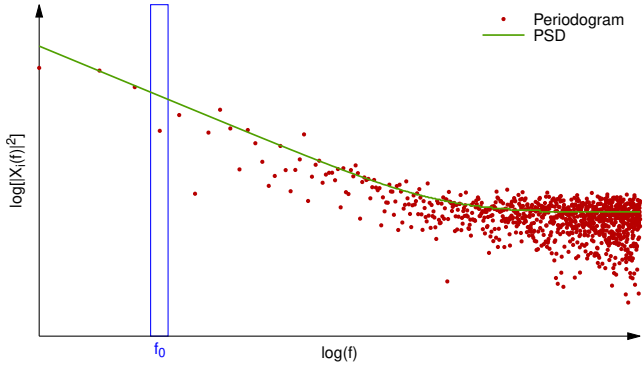}
    \caption{Periodogram of $x$ (white noise plus red noise). The PSD is the expectation of the periodogram.}
    \label{fig:periodogram}
\end{figure}

First, let us recall some basics of frequency analysis. Using a data record of duration $T$ sampled at a suitable frequency, the periodogram is

\begin{equation}
P_x(f) = \frac{2}{T}|X(f)|^2, \qquad f > 0
\label{eq:periodogram}
\end{equation}
where the factor ``2'' is needed for energy conservation after deleting the negative frequencies. The expectation of the periodogram is the Power Spectral Density (PSD),

\begin{equation}
S_x(f) = \mathbb{E} \left[ \frac{2}{T}|X(f)|^2\right], \qquad f > 0.
\label{eq:PSD}
\end{equation}
Figure \ref{fig:periodogram} shows the periodogram and the PSD. We estimate the PSD as the average periodogram, with the ultimate goal of expecting the red noise parameters of $r$ out of the measurement noise $n$. Of course $r$ is the same for all instruments, while the $n_i$ are specific to the $i$-th instrument and its environnement.\\
The total duration of the experiment is the major problem, as the lowest frequency of interest sets $T$. In turn, a long $T$ goes with a small number $p$ of averages because the total duration of the experiment is $pT$. In this paper we focus on the slowest red noise phenomena, up to years, for which we have to set $p = 1$. In other words, the phenomena of interest are so slow that we cannot average on multiple acquisitions.

\subsection{Estimators\label{sec2.3}}

We are now focusing on one bin of the periodogram of a single simultaneous measurement with $q$ instruments, e.g. $f_0$ as represented on Fig. \ref{fig:periodogram}. Let us emphasize on the term periodogram which designates a unique realization of the red noise since all instruments observe this red noise realization at the same time. Nevertheless, taking into account the uncorrelated white noises coming from the instruments, we have to deal with the PSD $S$. One bin of $S$ represents the power in a given bandwidth, i.e. the $2$-nd central moment, or variance. Hereinafter, we work on a generic bin, thus $S(f)$ at that frequency is replaced with $\sigma^2$.\\
Because the $N_i$ are all different, it is appropriate to use a weighted average, where the weights $\alpha_i$ are to be found for the optimum detection of $R$. We denote the \textbf{estimates} with a ``\textit{hat}'', then

\begin{equation}
    \widehat{\mu} = \frac{\sum_i^q \alpha_i X_i}{\sum_i^q \alpha_i},
    \label{eq:weighted_measurements}
\end{equation}
where $q$ is the number of instruments. The variance of the estimate $\widehat{\mu}$ is

\begin{equation}
\mathbb{V}\left[\widehat{\mu}\right] = \frac{\sum_i^q \alpha_i^2 \left(\sigma_{N,i}^2+\sigma_R^2\right)}{\left[\sum_i^q \alpha_i\right]^2}.
\label{eq:weight_variance}
\end{equation}
An optimal choice is obtained by solving,

\begin{equation}
\frac{\partial \mathbb{V}\left[\widehat{\mu}\right]}{\partial \alpha_i} = 0
\label{eq:gradient}
\end{equation}
which leads to the solution,

\begin{equation}
\alpha_i = \frac{1}{\sigma_{N,i}^2}.
\label{eq:alpha}
\end{equation}
Therefore the inverse-variance weighted average, described in \cite{hartung2008} with applications examples, has the least variance among all weighted averages. Then Eq. (\ref{eq:weight_variance}) becomes

\begin{equation}
\sigma_{\mu}^2 = \mathbb{V}\left[\widehat{\mu}\right] = \left( \sum_i^q \frac{1}{\sigma_{N,i}^2}\right)^{-1}.
\label{eq:weighted_noise_variance}
\end{equation}
Let us define now the two \textbf{estimators} of interest: the spectrum average weighted by the noise variance $\sigma_{N,i}^2$ and the cross-spectrum,

\begin{equation}
\begin{array}{l}
\widehat{S_{\mathrm{sa}}}=\displaystyle\left\{\Re\left[\sigma_{\mu}^2\sum_i^q\frac{X_i}{\sigma_{N,i}^2}\right]\right\}^2+\left\{\Im\left[\sigma_{\mu}^2\sum_i^q\frac{X_i}{\sigma_{N,i}^2}\right]\right\}^2 \\[0.3cm]
\widehat{S_{\mathrm{cs}}}=\langle \Re\left[ X_i\cdot\tilde{X}_j \right] \rangle_m \qquad \textrm{with} \quad i \ne j.
\end{array}
\label{eq:biased_estimator}
\end{equation}
Moreover $\sigma_{\mu}^2$ corresponds to the noise weight normalization factor defined in Eq. (\ref{eq:weighted_noise_variance}). Finally $\langle \cdot \rangle$ stands for the $m$ average over the different combinations of instruments with $m = \dbinom{q}{2}$ and $\tilde{\cdot}$ stands for the complex conjugate of the quantity which is below. Furthermore we have omitted in Eq. (\ref{eq:biased_estimator}) the measurement time factor $\frac{2}{T}$ which is necessary to have the dimension of a power per frequency for a better readability thereafter. In addition, only the random part has a direct influence on the probability density function. Denoting $\mathbb{E}[\cdot]$ the mathematical expectation of the quantity within the brackets,

\begin{equation}
\left\{\begin{array}{lcl}
\mathbb{E}\left[\widehat{S_{\mathrm{sa}}}\right]&=&\sigma_R^2 + \sigma_{\mu}^2\\[0.3cm]
\mathbb{E}\left[\widehat{S_{\mathrm{cs}}}\right]&=&\sigma_R^2
\end{array}\right.
\label{eq:expectation}
\end{equation}
which means that the spectrum average estimator is \textbf{biased}. Usually one removes the bias to have the s.a estimate average over realizations which tends towards the sought signal level $\sigma_R^2$. This gives a clear advantage to the c-s estimator. However, we will see that the computation of the confidence interval over the signal level $\sigma_R^2$ requires an estimation of this bias $\sigma_\mu^2$ whatever the chosen estimator, s.a or c-s. Therefore we want to estimate the PSD and we assume it follows a $1/f^\alpha$ power law, then we only have to estimate a level and exponent of the first frequency bins.\\
We now compare the estimator defined in Eq. (\ref{eq:biased_estimator}) by determining their variance. We can demonstrate provided that $\forall i,\ \sigma_{N,i}^2 = \sigma_N^2$ (see Annexe \ref{appendix:chap1}),

\begin{equation}
\mathbb{V}\left[\widehat{S_{\mathrm{cs}}}\right]\approx\left\{\begin{array}{lcl}
\mathbb{V}\left[\widehat{S_{\mathrm{sa}}}\right] & \textrm{if} & \sigma_R^2 \gg \sigma_N^2 \\[0.3cm]
\frac{q}{q-1}\mathbb{V}\left[\widehat{S_{\mathrm{sa}}}\right] & \textrm{if} & \sigma_R^2 \ll \sigma_N^2.
\end{array}\right.
\end{equation}
This is confirmed by Fig. \ref{fig:estimate_variance} which exhibits the variance of the estimates of both estimators applied to a signal composed of a mixture of uncorrelated white noise of level 1 arbitrary unit (a.u.) and a common $f^{-4}$ noise of level 4096 a.u. for 2 instruments. Therefore the variance decreases in $f^{-8}$ and Fig. \ref{fig:estimate_variance} compares these variances to the square of the PDF. At $f = 4$ a.u., the signal PSD is 16 times higher than the white level and therefore its square is 256 times higher. In this case, the variances of both estimators coincide. On the other hand, for frequencies higher than 16 a.u., the signal PSD is less than 16 times lower than the white level (256 for their squares) and the variance of the c-s estimates is 2 times higher than the variance of the s.a estimates. This seems to indicate a better efficiency of the s.a estimator. Indeed the spectrum average estimator is a sufficient estimator which means of minimal variance.

\begin{figure}[H]
\includegraphics[width=0.5\textwidth]{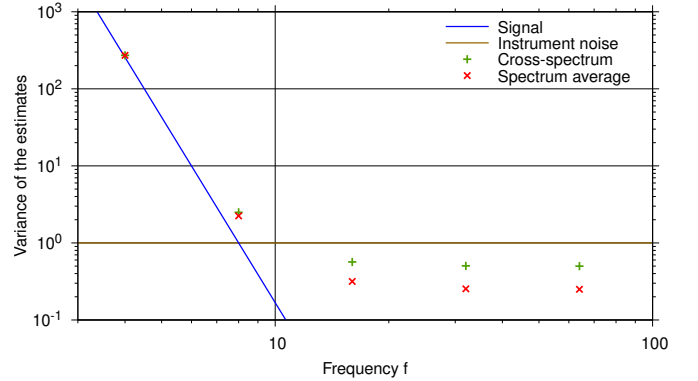}
\caption{Variance of the estimate with the signal variance which is of the form $(\sigma_R^2/f^{\alpha})^2$, where $\sigma_R^2=4096$ a.u. is the signal level and $\alpha = 4$ the red noise exponent. The noise model is a white noise of level $\sigma_N^2=1$ a.u. with 2 instruments.}
\label{fig:estimate_variance}
\end{figure}
 However what about the pdf of the estimates knowing the parameter $\sigma_R^2$ for a given frequency?

\section{Probability Density Function\label{sec3}}
\subsection{Spectrum Average Method\label{sec3.1}}

The spectrum average estimator leads to the following $\chi^2$ distribution with 2 degrees of freedom resulting from the real and imaginary part of the spectrum,

\begin{equation}
    p(\widehat{S_{\mathrm{sa}}}|\sigma_R^2) = \frac{e^{-\frac{\widehat{S_{\mathrm{sa}}}}{2\sigma^2}}}{2\sigma^2}
    \label{eq:av_PDF}
\end{equation}
where,
\begin{equation}
    \begin{array}{lcl}
        \sigma^2 &=& \frac{1}{2} \left(\sigma_\mu^2 + \sigma_R^2\right).
        \label{eq:Sav_Var}
    \end{array}
\end{equation}
where $\sigma_\mu^2$ is the weighted noise level according to Eq. (\ref{eq:weighted_noise_variance}) and $\sigma_R^2$ the signal level of interest.

\subsection{Karhunen-Lo\`{e}ve Transform\label{sec3.2}}

The KLT method, denoting to the Karhunen-Lo\`{e}ve transform, has been developed in \cite{lantz2019}. It uses the statistics of the data themselves instead of the statistics of the estimates. This method has the advantage to combine linearly independent Gaussian estimates. Furthermore it also forms a sufficient statistics like the s.a method. It is based on determining the covariance matrix $M$ associated to the real or imaginary part of the measurement $X_i$ obtained by the $q$ instruments,

\begin{equation}
\left\{\begin{aligned}
M_{ii} &= \frac{1}{2} \left(\sigma_{N,i}^2 + \sigma_R^2\right)\\
M_{ij} &= \frac{1}{2} \sigma_R^2 \qquad \textrm{with} \quad i \ne j
\end{aligned}\right.
\label{eq:covmat}
\end{equation}
where the extra factors $1/2$ come from Eq (\ref{eq:variance_factor}). This covariance matrix has to be diagonalized and we denote the eigenvalues $\lambda_i$. Their associated normalized eigenvectors are $V_i$ and the pdf is then given by

\begin{equation}
p(\widehat{S_{\mathrm{KLT}}}|\sigma_R^2) = \prod_{i=1}^{q}\frac{1}{\left(2\pi\lambda_i\right)^{\nu/2}}\e^{\left(-\frac{\sum_{j=1}^\nu w_{ij}^2}{2\lambda_i}\right)}
\label{eq:KLT_PDF}
\end{equation}
where $j$ highlights the real and imaginary part obtained through the Fourier transform therefore $\nu = 2$. Let us remind that $X$ corresponds to the matrix containing the set of Fourier transform of the measurements at the output of each instrument. The numerator of the exponential argument is then

\begin{equation}
w = X \cdot V
\label{eq:w}
\end{equation}
where $V$ are the eigenvectors obtained from the diagonalized covariance matrix.

\subsection{Cross-spectrum\label{sec3.3}}

The cross-spectrum estimator leads to the variance-gamma (V$\Gamma$) distribution for 2 instruments as described in Section III in \cite{baudiquez2020} but for more than 2 instruments it is no longer the case. Having no exact solution known nowadays, we give an approximation of it. The process is the same until the establishment of the $\chi^2$ linear combination. First we perform an orthonormalization by using the Householder transformation to define a basis of unit vectors that are orthogonal to each other. We define $\mathcal{W}$ the matrix where each column contains the standard deviation of the spectrum according to Eq. (\ref{eq:time-freq}) as

\begin{equation}
\mathcal{W} = \frac{1}{\sqrt{2}}
\begin{pmatrix}
\sigma_{N,1} & 0 & \ldots & \ldots & 0\\
0 & \sigma_{N,2} & 0 & \ldots & 0\\
0 & 0 &\sigma_{N,3} & \ldots & 0\\
\vdots & \vdots & \vdots & \ddots & \sigma_{N,q} \\
\sigma_R & \sigma_R & \sigma_R & \ldots & \sigma_R
\end{pmatrix}.
\label{W_matrix}
\end{equation}
All the measurement noises are independent, as assumed, whereas the signal is common. Then $\mathcal{W}$ is projected onto the orthogonal basis and we compute the eigenvalues $\lambda_j$ of the resulting components. This leads to a linear combination of $\chi^2$ distribution as follows,

\begin{equation}
    \widehat{S_{\mathrm{cs}}} = \sum_j^q \lambda_j \chi_k^2
    \label{chi2_linear_combination}
\end{equation}
where $k$ is the number of degrees of freedom corresponding to each eigenvalue, e.g. equal to 2 for the real and imaginary part without degeneration. We respectively used the DGEQRF and DSYEV LAPACK subroutine to perform the orthonormalization and compute the eigenvalues. In the special case of 2 instruments we obtain the subtraction of two $\chi^2$ random variables with the same number of degrees of freedom. The characteristic function of the $\chi_k^2$ distribution is defined as

\begin{equation}
    \phi_j(t) = (1-2i\lambda_jt)^{-k/2}
    \label{characteristic function}
\end{equation}
where $i$ is the imaginary unit and we apply a variable change of $-t$ for the negative eigenvalues. The $\chi^2$ distributions according to Eq. (\ref{chi2_linear_combination}) being independent, the characteristic function of the c-s becomes

\begin{equation}
    \phi(t) = \prod_j^q \phi_j(t).
    \label{chi2_product}
\end{equation}
It leads to the moment generating function of the V$\Gamma$ distribution for 2 instruments but it is no longer the case for more instruments. When all the instruments have the same level of intrinsic noise $\sigma_\mathrm{n}^2$, the diagonalization of the matrix $\mathcal{W}$ defined by Eq. (\ref{W_matrix}) leads to two eigenvalues. One is unique and the second one has a degeneration of $q-1$ with $q$ the number of instruments. Consequently, it leads to the difference of two $\chi^2$ random variables with different degrees of freedom. However even if it looks like the case with 2 instruments, the difference in the degrees of freedom of the $\chi^2$ distributions has no analytical solution. Therefore the probability density function of the c-s for any noise level is defined as

\begin{equation}
    p(\widehat{S_{\mathrm{cs}}}|\sigma_R^2) = \frac{1}{2\pi}\int_\mathbb{R} \e^{-it\widehat{S_{\mathrm{cs}}}}\phi(t)dt.
    \label{eq:cs_PDF}
\end{equation}
We perform the integration by using the Simpson method only on the positive side because the real part of this function is even whereas the imaginary part is odd. Figure \ref{fig:cs_pdf} shows that the theoretical probability density function fits very well the histogram obtained by $10^7$ Monte Carlo simulations for 5 instruments. The variance of each white noise is the same $\sigma_N^2 = 10$ a.u. whereas the signal level is $\sigma_R^2 = 6$ a.u.

\begin{figure}
\includegraphics[width=0.5\textwidth]{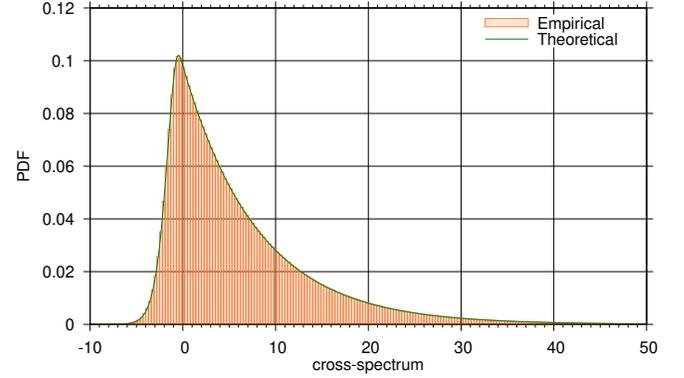}
\caption{Comparison of the empirical (red boxes) and theoretical (green line) pdf of the c-s for 5 instruments where the variances are $\sigma_R^2 = 6$ a.u. and $\sigma_N^2 = 10$ a.u.}
\label{fig:cs_pdf}
\end{figure}

\subsection{Bayesian inference\label{sec3.4}}
\subsubsection{A posteriori distribution\label{sec3.4.1}}

We seek to determine a confidence interval on $\sigma_R^2$, but Eq. (\ref{eq:av_PDF}), (\ref{eq:KLT_PDF}) and (\ref{eq:cs_PDF}) define the pdf of a set of measurement $X$ given the sought parameter $\sigma_R^2$. So we have to solve the inverse problem which means to determine the pdf of $\sigma_R^2$ given a set of measurement $X$ called the \textit{posterior} distribution. The Bayes theorem leads to the following relation,

\begin{equation}
\left\{\begin{array}{l}
p(\sigma_R^2|X) \propto p(X|\sigma_R^2) \cdot \pi(\sigma_R^2)\\[0.3cm]
\int_0^\infty p(\sigma_R^2|X) d\sigma_R^2 = 1
\end{array}\right.
\label{eq:posterior}
\end{equation}
where $\pi(\sigma_R^2)$ is the prior, i.e. the pdf before any measurement. One of the main issue of Bayesian analysis concerns the choice of this prior.

\subsubsection{Choice of the prior\label{sec3.4.2}}

In order to be as general as possible, we will assume a total ignorance of the signal level. In such a case, it is generally considered that any order of magnitude has the same probability which suggests a constant prior in a logarithmic scale, i.e $\pi\left(\sigma_R^2\right)=1/\sigma_R^2$. However, our perfect knowledge of the noise level induces an implicit scale factor. In other words, since we did not remove the "bias" $\sigma_\mu^2$ in Eq. (\ref{eq:expectation}), the s.a estimator is shifted by $\sigma_\mu^2$. In a very similar case \cite{mchugh96b}, we decided that the true parameter should be the sum of both levels $\theta=\sigma_\mu^2+\sigma_R^2$. Moreover according to Eq. (\ref{eq:weighted_noise_variance}) higher noise will have lower weight and in our case, since the mathematical expectation of the s.a estimator is $\sigma_\mu^2+\sigma_R^2$, it comes naturally that the true parameter should be:

\begin{equation}
\theta=\sigma_\mu^2+\sigma_R^2.
\label{eq:prior_parameter}
\end{equation}
From these considerations, we will choose $\pi(\theta)=1/\theta=\frac{1}{\sigma_\mu^2+\sigma_R^2}$ and then, our prior for the s.a estimator will be
\begin{equation}
\pi(\sigma_R^2)\propto \frac{1}{\sigma_\mu^2+\sigma_R^2}.\label{eq:prior}
\end{equation}
In order to be fair in the trial of c-s against s.a, the same prior will be used for both estimators.\\
In the following we will compare the different methods, starting with the spectrum average and KLT in Sec. \ref{sec4}.

\section{Spectrum average and KLT comparison\label{sec4}}
\subsection{A particular case: all the instruments have the same variance\label{sec4.1}}

Let us define $\forall i, \sigma_{N,i}^2=\sigma_N^2$, i.e. all the $q$ instruments have the same noise level. At a first step we determine the s.a pdf, in this case according to Eq. (\ref{eq:weighted_noise_variance}) and (\ref{eq:variance_factor}), the variance defined by Eq. (\ref{eq:Sav_Var}) leads to the following expression,

\begin{equation}
\sigma^2 = \displaystyle \frac{1}{2}\left(\frac{\sigma_N^2}{q} + \sigma_R^2\right).
\end{equation}
From Eq. (\ref{eq:biased_estimator}), the estimate $\widehat{S_{\mathrm{sa}}}$ now becomes
\begin{equation}
    \begin{array}{lcl}
        \widehat{S_{\mathrm{sa}}} &=& \displaystyle \left\{\Re\left[\sigma_\mu^2\sum_i^q\frac{X_i}{\sigma_{N,i}^2}\right]\right\}^2+\left\{\Im\left[\sigma_\mu^2\sum_i^q\frac{X_i}{\sigma_{N,i}^2}\right]\right\}^2\\[0.3cm]
        &=& \displaystyle \frac{1}{q^2}\left(\left\{\Re\left[\sum_i^q X_i\right]\right\}^2+\left\{\Im\left[\sum_i^q X_i\right]\right\}^2\right)
        \label{eq:av_numerator}
    \end{array}
\end{equation}
According to Eq. (\ref{eq:av_PDF}), the s.a pdf is given by
\begin{equation}
    p(\widehat{S_{\mathrm{sa}}}|\sigma_R^2) = \displaystyle \frac{e^{-\frac{\frac{1}{q^2}\left\{\Re\left[\sum_i^q X_i\right]^2+\Im\left[\sum_i^q X_i\right]^2\right\}}{\frac{\sigma_N^2}{q}+\sigma_R^2}}}{\frac{\sigma_N^2}{q}+\sigma_R^2}.
    \label{eq:av_PDF_same_noise}
\end{equation}
In a second step let us define the KLT pdf. The eigenvalues of the covariance matrix resulting from Eq. (\ref{eq:covmat}) are given by

\begin{equation}
    \begin{array}{lcl}
        \lambda_1 &=& \frac{1}{2}\left(\sigma_N^2 + q\sigma_R^2\right)\\[0.3cm]
        \lambda_i &=& \frac{1}{2}\sigma_N^2 \qquad \textrm{with} \quad i \in \left\{2,...,q\right\}
        \label{eq:eigenvalues}
    \end{array}
\end{equation}
The first and highest eigenvalue being the only one to depend of $\sigma_R^2$, we solely define its associated eigenvector
\begin{equation}
V_1 = \frac{J_{q,1}}{\sqrt{q}}
\end{equation}
where $J_{q,1}$ is the all-ones column vector. Then the numerator in the exponential in Eq. ({\ref{eq:KLT_PDF}}) is
\begin{equation}
    \begin{array}{lcl}
        \sum_j^\nu \hat{w}_{1,j}^2 &=& \sum_j^\nu \left[X_j \cdot V_1\right]^2\\[0.3cm]
        &=& \frac{1}{q}\sum_j^\nu \left[X_j \cdot J_{q,1}\right]^2\\[0.3cm]
        &=& \frac{1}{q}\sum_j^\nu \left[\sum_i^q X_{ij}\right]^2\\[0.3cm]
        &=& \frac{1}{q} \left\{\Re\left[\sum_i^q X_i\right]^2+\Im\left[\sum_i^q X_i\right]^2\right\}.
        \label{eq:cs_numerator}
    \end{array}
\end{equation}
The KLT pdf defined by Eq. (\ref{eq:KLT_PDF}) is given by
\begin{equation}
    p(\widehat{S_{\mathrm{KLT}}}|\sigma_R^2) = C \frac{e^{-\frac{\frac{1}{q}\left\{\Re\left[\sum_i^q X_i\right]^2+\Im\left[\sum_i^q X_i\right]^2\right\}}{\sigma_N^2+q\sigma_R^2}}}{\pi\left(\sigma_N^2+q\sigma_R^2\right)}
    \label{eq:cs_PDF_same_noise}
\end{equation}
where $C$ is the Gaussian remaining product with a variance depending only on the measurement noise level. However what we want to characterize is not the estimates but the parameter $\sigma_R^2$. According to Eq. (\ref{eq:posterior}), the pdf of the true parameter $\sigma_R^2$ is proportional to the prior $\pi(\sigma_R^2)$ multiplied respectively by Eq. (\ref{eq:av_PDF_same_noise}) and (\ref{eq:cs_PDF_same_noise}) for the s.a and KLT estimates. The Bayes theorem leads then to

\begin{equation}
    \begin{array}{lcl}
        p(\sigma_R^2|\widehat{S_{\mathrm{sa}}}) &\propto& \pi(\sigma_R^2) \displaystyle\frac{e^{-\frac{\frac{1}{q}\left\{\Re\left[\sum_i^q X_i\right]^2+\Im\left[\sum_i^q X_i\right]^2\right\}}{\sigma_N^2+q\sigma_R^2}}}{\sigma_N^2+q\sigma_R^2}\\[0.3cm]
        \label{eq:av_posterior}
    \end{array}
\end{equation}
and
\begin{equation}
    \begin{array}{lcl}
        p(\sigma_R^2|\widehat{S_{\mathrm{KLT}}}) &\propto& \pi(\sigma_R^2) \displaystyle\frac{e^{-\frac{\frac{1}{q}\left\{\Re\left[\sum_i^q X_i\right]^2+\Im\left[\sum_i^q X_i\right]^2\right\}}{\sigma_N^2+q\sigma_R^2}}}{\sigma_N^2+q\sigma_R^2}.\\[0.3cm]
        \label{eq:klt_posterior}
    \end{array}
\end{equation}
Multiplying respectively Eq. (\ref{eq:av_posterior}) and (\ref{eq:klt_posterior}) by a factor $1/q$ and $\pi$ does not change the pdf since it is normalized. It is exactly the same for Eq. (\ref{eq:klt_posterior}) where $C$ does not depend on $\sigma_R^2$ and vanish through the normalization. Therefore both expressions are exactly the same. It should also be noted that the noise level $\sigma_N^2$ is necessary in both cases and the bias does not influence the sought parameter density whereas it does regarding the estimates. This implies a very interesting consequence: both pdf for the s.a and KLT leads to the exact \textbf{same} confidence interval for the same noise level.

\subsection{General case\label{sec4.2}}

In this part any number of instruments and different noise level for each of them can be considered. In Section \ref{sec4.1}, we showed analytically that both methods lead to the same pdf of the signal level knowing the estimates in the event that all noise levels are the same. However when each noise level is different Eq. (\ref{eq:eigenvalues}) giving the relation between the eigenvalues and the signal becomes much more complicated without degeneration. In this case, let us consider a number of instruments solely up to 5, refering as instance to the number of radio telescopes (RTs) part of the LEAP project. Then we make several empirical comparisons by computing the upper limit at 95\% for the spectrum average and KLT methods. It should be noticed that the 5\% lower bound has no interest since we are more particularly interested in the case where the signal is weaker than the noise level. This bound then greatly depends on the prior and is very close to zero.\\
Table \ref{table:diff_RT_average} gives the average over $1\,000$ realizations of the 95\% upper bound for 2 to 5 RTs. The signal and noise levels are respectively $\sigma_R^2 = 1$ a.u. and $\sigma_{N,i}^2 = i$ a.u. where $i$ is the $i$-th RT. Then the $2$-nd and $3$-rd RT are respectively $2$ and $3$ times more noisy than the first one and so forth.\\
First, these comparisons show as expected that the 95\% bounds obtained by both estimators as in Sec. \ref{sec4.1} for the same noise variance, are exactly the same.\\
Second, the mean and median are decreasing as the number of RTs increases. As a consequence adding measuring instruments or RTs always add information about the signal level or in the worst case is useless but never worsen it. On the other side the upper bound maximum values obtained depend strongly on the stochastic behavior of the measurements.\\
Finally, it should be noticed that both methods require the noise level knowledge for the expression of the probability density function. The spectrum average method being the fastest way to compute the confidence interval is then to be privileged. Therefore we will only compare the spectrum average method with the cross-spectrum in the next section.

\begin{table}
    \caption{Upper limit average of the parameter $\sigma_R^2$ taking into account $2$ to $5$ RTs. These data were obtained from a set of $1\,000$ simulated spectra. The signal and noise level used for the computation are $\sigma_R^2 = 1$ and $\sigma_{N,i}^2 = i$ where $i$ is the index of the RT.\label{table:diff_RT_average}}
\begin{center}
\begin{tabular}{|c|c|c|c|c|c|}
\hline
\multirow{2}{*}{} & \multicolumn{5}{c|}{Spectrum average / KLT \quad 95\% upper limit} \\
\cline{2-6}
RTs number & Mean & Median & Std & Min & Max \\
\hline
\hline
2 & 17.44 & 12.88 & 3.10 & 6.30 & 115.32 \\
3 & 16.32 & 11.78 & 2.39 & 5.16 & 91.78 \\
4 & 15.66 & 11.10 & 2.95 & 4.54 & 108.82 \\
5 & 14.84 & 10.67 & 2.28 & 4.14 & 86.99 \\
\hline
\end{tabular}
\end{center}
\end{table}

\section{95\% upper limit: spectrum average vs cross-spectrum\label{sec5}}

We have set the direct problem, i.e. the statistics of the s.a or c-s knowing the signal level and noise level (which is assumed to be known), respectively in Sections \ref{sec3.1} and \ref{sec3.3}. Now we tackle the inverse problem from the direct problem, i.e. the statistics of the signal level knowing the s.a or c-s estimate. The Bayes theorem enables us to establish this link as described in section \ref{sec3.4}. The posterior distribution of the s.a and c-s are given by

\begin{equation}
    \begin{array}{lcl}
        p(\sigma_R^2|\widehat{S_{\mathrm{sa}}}) &\propto& \frac{1}{(\sigma_\mu^2+\sigma_R^2)^2} \e^{\frac{-\widehat{S_{\mathrm{sa}}}}{\sigma_\mu^2+\sigma_R^2}}
        \label{eq:sa_posterior}
    \end{array}
\end{equation}
and
\begin{equation}
    \begin{array}{lcl}
        p(\sigma_R^2|\widehat{S_{\mathrm{cs}}}) &\propto& \frac{1}{2\pi(\sigma_\mu^2+\sigma_R^2)}\int_\mathbb{R} \e^{-it\widehat{S_{\mathrm{cs}}}}\phi(t)dt
        \label{eq:cs_posterior}
    \end{array}
\end{equation}
where $\sigma_\mu^2$ is the noise variance weighting according to Eq. (\ref{eq:weighted_noise_variance}). Let us describe our simulation algorithm in order to assess the 95\% upper limit.\\
First simulation ($S_1$ to $S_3$): simulate a set of real data from $q$ instruments, assuming the red noise level is known (as well as, of course, the measurement noise levels).
\begin{enumerate}[label=\subscript{S}{{\arabic*}}:]
    \item Assign the number of RTs, the noise variance of each one and the sought \textbf{true} signal level.
    \item Generate a set of spectral measurement according to Eq. (\ref{eq:time-freq})
    \item Compute the s.a and c-s estimates, as stated in Eq. (\ref{eq:biased_estimator}), which are now fixed as parameters.
\end{enumerate}
Second simulation: we no longer modify the data (these are acquired measurement results) and we look for a confidence interval on the red noise, assuming the level of the measurement noise is known.
\begin{enumerate}[label=\subscript{S}{{\arabic*}}:]
    \setcounter{enumi}{3}
    \item Define any basis and perform an orthogonalization and normalization of it by using the DGEQRF subroutine from LAPACK
    \item Establish, from Eq. (\ref{W_matrix}), one $\mathcal{W}$ matrix for each signal level \textbf{varying} from $0$ to an upper limit for which Eq. (\ref{eq:sa_posterior}) and (\ref{eq:cs_posterior}) are close enough to zero according to the required precision.
    \item Peform $S_7$ to $S_{11}$ for each $\sigma_R^2$ value.
    \item Project the $\mathcal{W}$ matrix onto the orthogonal basis.
    \item Compute the c-s denoted $\mathcal{Z}$ from the result of $S_6$.
    \item Determine the eigenvalues of $\mathcal{Z}$ by using the DSYEV subroutine from LAPACK which has now the form of Eq. (\ref{chi2_linear_combination}).
    \item Define the product of each characteristic function defined by Eq. (\ref{characteristic function}).
    \item Compute the posterior distribution respectively of the s.a and c-s estimates according to Eq. (\ref{eq:sa_posterior}) and (\ref{eq:cs_posterior}). For the c-s, we perform a numerical integration of one signal value by using the Simpson method.
    \item Normalize the s.a and c-s posterior pdf.
    \item Determine the cumulative distribution function (cdf) by integrating the s.a and c-s posterior pdf and find the 95\% upper limit corresponding onto the cdf value associated to the signal level.
\end{enumerate}
The loops for the different values of the signal are computed in parallel in order to save computing time. Let us give an example of such a process. We set the number of RTs to 5 and the variances of the signal and noise are respectively $\sigma_R^2 = 6$ a.u., $\sigma_\mathrm{n}^2 = 10$ a.u. Then we produce 2 sets of random measurement with these parameters, shown in Table \ref{table1}. The first measurement set gives respectively $\widehat{S_{\mathrm{sa},1}} = 14.886$ a.u. and $\widehat{S_{\mathrm{cs},1}} = 13.226$ a.u. for the s.a and c-s estimates whereas the second one gives $\widehat{S_{\mathrm{sa},2}} = 20.730$ a.u. and $\widehat{S_{\mathrm{cs},2}} = 18.564$ a.u. It leads for the first set to the 95\% upper limit on the signal $\sigma_R^2$ following value, $125.8$ for the s.a and $127.3$ for the c-s. Furthermore the second set gives us $167.1$ for the s.a and $164.8$ for the c-s. These results show that either the c-s or the s.a can be the most efficient even with the same parameters, then it only depends on the measurement set. However, the difference between the 95\% upper limit for both methods is relatively low.

\begin{table}
    \caption{Measurement set for the outputs of each RT (5 in total) where $\sigma_R^2 = 6$ a.u. and $\sigma_N^2 = 10$ a.u.\label{table1}}
\begin{center}
\begin{tabular}{|c|c|c|c|c|}
\hline
\multirow{2}{*}{} & \multicolumn{2}{c|}{measurement set 1} & \multicolumn{2}{c|}{measurement set 2} \\
\cline{2-5}
& Real part & Imaginary part & Real part & Imaginary part\\
\hline
\hline
$X_1$  & -3.8947 & -1.7994 & -0.1494 & 8.9456 \\
$X_2$  & -5.0950 & -3.9125 & -0.5275 & 4.4659 \\
$X_3$  & -2.5133  & -5.5431 &  0.2176 & 5.7742 \\
$X_4$  & 0.6433  & -1.9566 & 1.6044  & 3.2146 \\
$X_5$  & -0.2294 & -2.5738 & -0.5284 & 0.3563 \\
\hline
\end{tabular}
\end{center}
\end{table}

Let us now compare the s.a and c-s 95\% upper limit over $100$ simulations as shown in Table \ref{table_upper_limit} for the sought signal level set to $6$ a.u. and a noise level equal to $10$ a.u. for each RT. The 95\% upper limit is given respectively for, from the top of the Table to the bottom,  the spectrum average, the cross-spectrum and the ratio of the 95\% bound of s.a over c-s. The mean and median are decreasing when the number of RTs is increasing. However for 4 RTs the results are much more lower but it is just an artefact of ``luck". Indeed the maximum value is $1.4$ times lower than for 5 RTs and the standard deviation (std) is also very much more lower. The sample size can have a significant effect on the values obtained but is necessary to have a good precision with a reasonable computation time. However, the minimum value of the 95\% bound obtained for both methods permits to override this randomness. Indeed when the cross-spectrum estimate is negative or the spectrum average estimate tends towards zero it leads to the smallest 95\% bound. Whereas the maximum 95\% bound obtainable for a reasonable amount of simulations can ``wriggle" a lot as the tail of the posterior pdf is very long especially with higher noise level than signal level which is of interest. The minimum value shows as expected an improvement with the increase in number of RTs. It seems that the s.a method gives the most stringent confidence interval.\\
Figure \ref{fig:95_hist} shows the histogram of the 95\% limit with $5$ RTs for $10\, 000$ realizations, $\sigma_R^2 = 6$ a.u. and $\sigma_N^2 = 10$ a.u. Both histograms exhibit a similar distribution which extend up to high values. However the first bin corresponding to the lowest 95\% bound shows a high number of realizations for the c-s method. This can be explained by a negative estimate for the cross-spectrum which may corresponds to a spectrum average estimate having a not so small value and so a higher 95\% bound. Figure \ref{fig:cs_vs_sa_upper_bound} shows the comparison of the 95\% upper limit for the s.a and c-s methods for a window of hundred data among the same set of realizations. The $6\, 620$-th realization framed by a blue rectangle highlights the fact that the c-s can sometimes be much more stringent than the s.a method. However in most of the other realizations we notice that the 95\% limit is almost the same.\\
Figure \ref{fig:95_ratio_cs_vs_sa} depict the 95\% upper bound median among $1\, 000$ simulations with $5$ RTs, for the s.a over c-s ratio depending on the signal-to-noise level ratio (with $\sigma_N^2 = 1$ a.u.). When $\sigma_R^2 \ll \sigma_N^2$ then the s.a seems to be the most stringent most of the time. However when the signal level becomes higher than the noise level, both the s.a and the c-s methods give in median the same 95\% limit.\\
 Considering all these observations it is wiser to compute both estimators and use the most restrictive one. Even if most of the time both estimators give a very close upper bound, sometimes the gap is clearly significant.

\begin{table}
    \caption{95\% upper limit statistics for the s.a (top), c-s (middle) and the ratio of the s.a by the c-s over $100$ simulations where $\sigma_R^2 = 6$ a.u. and $\sigma_N^2 = 10$ a.u. Each rows respectively from the left to the right corresponds to the number of RTs, the mean, median, standard deviation, minimun and maximum value of the 95\% upper bound.\label{table_upper_limit}}
\begin{center}
\begin{tabular}{|c|c|c|c|c|c|}
\hline
\multirow{2}{*}{} & \multicolumn{5}{c|}{Spectrum average 95\% upper limit} \\
\cline{2-6}
RTs number & Mean & Median & Std & Min & Max \\
\hline
\hline
2 & 112.99 & 79.45 & 32.93 & 48.50 & 440.60 \\
3 & 98.41 & 72.60 & 35.66 & 31.70 & 453.20 \\
4 & 78.00 & 51.50 & 18.30 & 23.80 & 260.10 \\
5 & 90.11 & 67.95 & 28.47 & 19.00 & 373.40 \\
\hline
\hline
\multirow{2}{*}{} & \multicolumn{5}{c|}{Cross-spectrum 95\% upper limit} \\
\cline{2-6}
RTs number & Mean & Median & Std & Min & Max \\
\hline
\hline
2 & 116.49 & 83.00 & 27.38 & 67.90 & 388.90 \\
3 & 99.74 & 79.65 & 34.54 & 41.00 & 443.40 \\
4 & 76.37 & 54.10 & 18.03 & 28.50 & 255.80 \\
5 & 91.87 & 65.35 & 28.98 & 22.20 & 380.20 \\
\hline
\hline
\multirow{2}{*}{} & \multicolumn{5}{c|}{s.a/c-s 95\% upper limit} \\
\cline{2-6}
RTs number & Mean & Median & Std & Min & Max \\
\hline
\hline
2 & 0.97 & 0.90 & 0.12 & 0.71 & 2.21 \\
3 & 0.98 & 0.94 & 8.13 $\times 10^{-2}$ & 0.74 & 1.79 \\
4 & 1.02 & 0.98 & 4.86 $\times 10^{-2}$ & 0.74 & 1.50 \\
5 & 0.97 & 0.96 & 3.73 $\times 10^{-2}$ & 0.78 & 1.34 \\
\hline
\end{tabular}
\end{center}
\end{table}

\begin{figure}
    \includegraphics[width=\linewidth]{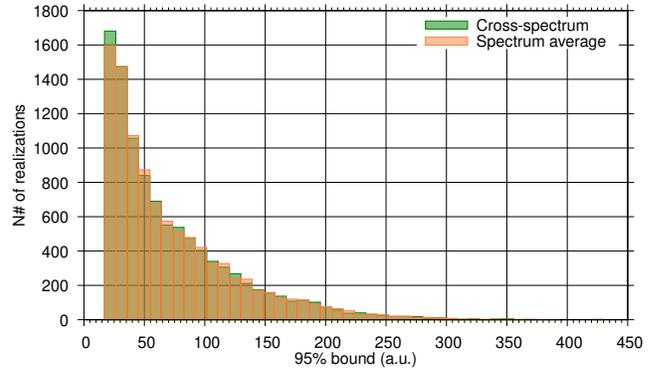}
    \caption{Histogram comparison of 95\% upper bound between the c-s and s.a for $10\, 000$ realizations. The parameters are set for 5 RTs, $\sigma_R^2 = 6$ a.u. and $\sigma_N^2 = 10$ a.u.}
    \label{fig:95_hist}
\end{figure}

\begin{figure}
\includegraphics[width=0.5\textwidth]{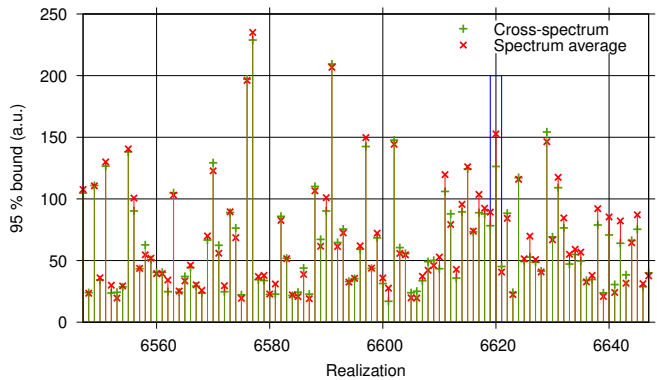}
\caption{Set of $100$ realizations for 5 RTs of 95\% bounds for cross-spectrum (green $+$) and spectrum average (red $\times$) where $\sigma_R^2 = 6$ a.u. and $\sigma_N^2 = 10$ a.u. }
\label{fig:cs_vs_sa_upper_bound}
\end{figure}

\begin{figure}
    \includegraphics[width=\linewidth]{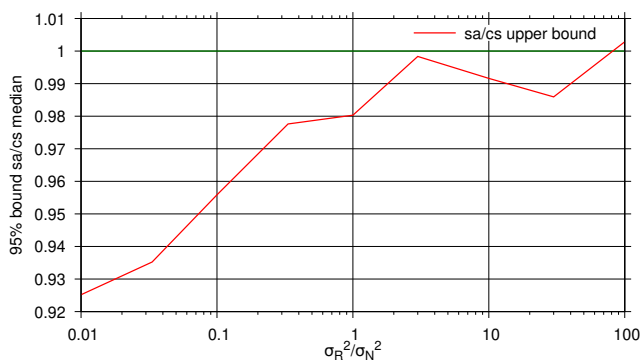}
    \caption{Evolution of the 95\% upper bound median of the s.a over c-s ratio obtained for $1\, 000$ realizations depending on the signal-to-noise level ratio. The parameters are set for 5 RTs and $\sigma_N^2 = 1$ a.u. The red curve corresponds to the sa/cs upper bound and the green curve indicates when both s.a and c-s have the same upper bound in median.}
    \label{fig:95_ratio_cs_vs_sa}
\end{figure}

\section{Conclusion\label{sec6}}

First, we demonstrated that the spectrum average variance is $q/(q-1)$ lower than the cross-spectrum variance.\\

Second, in order to assess the confidence interval of the signal level we defined its probability density function knowing  the s.a and c-s estimates but also the noise of each instruments (radio telescopes). In addition a method directly using the statistics of the measurement (KLT) has also been compared. It turns out that the KLT and the s.a methods lead to the exact same pdf of the signal level $\sigma_R^2$ knowing the estimates, so the precision is the same. Furthermore whereas the exact density of the cross-spectrum leads to the V$\Gamma$ distribution for 2 instruments. it is no longer the case for more instruments. We proposed a generalized method which implies a numerical integration of the characteristic function product. This method works very well according to the Monte Carlo simulations.\\

Finally the efficiency of both estimators, the spectrum average versus the cross-spectrum, is highlighted through the comparison of the 95$\%$ Bayesian upper limit. We found a slight advantage for the spectrum average estimator when the noise level is higher than the signal level. However we showed that sometimes the c-s gives the most stringent confidence interval but above all a little more often than the s.a for the lowest upper limit. Nevertheless it is the s.a method which gives us the minimum 95\% limit reachable. To conclude it is wiser to compute both estimates and use the most stringent.

\section*{Acknowledgement\label{sec7}}

This work was partially funded by the ANR Programmes d'Investissement d'Avenir (PIA) Oscillateur IMP (Project 11-EQPX-0033) and FIRST-TF (Project 10-LABX-0048).

\bibliography{Ref-local}

\newpage
\appendix
    \subsection*{Glossary of symbols}
        \begin{center}
            \begin{tabular}{ll}
                \hline
                $q$ & Number of instruments\\
                \hline
                $r(t)$ & Common signal measured by $q$ RTs (red noise)\\
                \hline
                $R(f)$ & Fourier transform of $r(t)$\\
                \hline
                $S_r$(f) & Power spectral density of $r(t)$\\
                \hline
                $n_i(t)$ & Intrinsic white noise of the $i$-th RT\\
                \hline
                $N_i(f)$ & Fourier transform of $n_i(t)$\\
                \hline
                $S_{n,i}$(f) & Power spectral density of $n_i(t)$\\
                \hline
                $x_i(t)$ & $x_i(t) = r(t) + n_i(t)$, received at the output of\\
                & the $i$-th RT\\
                \hline
                $X_i(f)$ & Fourier transform of $x_i$(t)\\
                \hline
                $S_{x,i}$(f) & Power spectral density of $x_i(t)$\\
                \hline
                \multicolumn{2}{l}{$\widehat{}\ $ estimate as in $\widehat{S}$. Here we consider three estimators,}\\
                $\widehat{S_{\mathrm{sa}}}$ & Spectrum average\\
                $\widehat{S_{\mathrm{KLT}}}$ & Karhunen-Lo\`{e}ve transform\\
                $\widehat{S_{\mathrm{cs}}}$ & Cross-spectrum\\
                \hline
                $\sigma_R^2$ & Variance of $R$ in a bandwidth, i.e. the power in\\
                & one bin of $S(f)$. It takes three different flavors:\\
                & s.a, KLT or c-s\\
                \hline
                $\sigma_{N,i}^2$ & Same as above, with the noise of the $i$-th RT\\
                \hline
                $\sigma_\mu^2$ & Noise weight factor, inverse of the sum of the\\
                & inverse of $\sigma_{N,i}^2$\\
                \hline
            \end{tabular}
        \end{center}

    \subsection*{Variance of the estimators $\widehat{S_{\mathrm{sa}}}$ and $\widehat{S_{\mathrm{cs}}}$\label{appendix:chap1}}
        \subsubsection{Measurements}

    Let us define $q$ instruments measurements $X_1$, $X_2$, $\ldots$ and $X_q$ as
    $$
    X_j = N_j + i N_j' + R + i R'
    $$
    where $N_j, N_j'$ are independent Gaussian centered random variables of variance $\sigma_N^2/2$ and $S, S'$ are independent Gaussian centered random variables of variance $\sigma_R^2/2$.

    	\subsubsection{Estimators}
    The estimator $\widehat{S_{\mathrm{cs}}}$ is defined by Eq. \ref{eq:biased_estimator} as
    \begin{equation}
        \begin{aligned}
            \widehat{S_{\mathrm{cs}}} = \displaystyle \frac{1}{\binom{q}{2}}\sum_{j=1}^{q-1}\sum_{k=j+1}^{q} \Re [&(N_j+i N'_j + R + i R') {\times}\\
            & \quad {\times} (N_k - i N'_k + R - i R')].\label{eq:def_cs}
        \end{aligned}
    \end{equation}
    On the other hand, $\widehat{S_{\mathrm{sa}}}$ is defined by Eq. \ref{eq:biased_estimator} as
    \begin{equation}
    \widehat{S_{\mathrm{sa}}} = \left(\sum_j^q \frac{N_j+qR}{q}\right)^2+\left(\sum_j^q \frac{N_j'+qR'}{q}\right)^2.\label{eq:def_sa}
    \end{equation}

    	\subsubsection{Statistics reminder}
    If $A$ and $B$ are 2 independent random variables of zero expectation
    \begin{equation}
    \var[AB]=\var[A]\var[B]\label{eq:prop1}
    \end{equation}
    according to Eq. (a) from \cite{barnett1955} where $\var[\cdot]$ stands for the variance of the quantity within the brackets.
    Moreover according to the Isserlis' theorem \cite{isserlis1918},
    \begin{equation}
        \begin{aligned}
            \var[A^2] &=\mxp[A^4]-\left\{\mxp[A^2]\right\}^2 = 3\left\{\mxp[A^2]\right\}^2-\left\{\mxp[A^2]\right\}^2\\
            &=2\var^2[A]\label{eq:prop2}
        \end{aligned}
    \end{equation}
    where $\mxp[\cdot]$ stands for the mathematical expectation of the quantity within the brackets. It is also useful to consider the covariances. If $A, B, C, D$ are 4 Gaussian centered random variable
    \begin{equation}
    \mxp[ABCD]=\mxp[AB]\cdot\mxp[CD]+\mxp[AC]\cdot\mxp[BD]+\mxp[AD]\cdot\mxp[BC].
    \end{equation}

    \noindent
    If $A, B, C, D$ are 4 \textbf{independent} Gaussian centered random variables, this can be derived to the following particular cases (Isserlis' theorem \cite{isserlis1918}):
    \begin{itemize}
    	\item $\mxp[ABCD]=\mxp[AB]\cdot\mxp[CD]+\mxp[AC]\cdot\mxp[BD]+\mxp[AD]\cdot\mxp[BC]=0$ since each mathematical expectation product $\mxp[XY]$ is null
    	\item $\mxp[A^2BC]=\mxp[A^2]\cdot\mxp[BC]+2\mxp[AB]\cdot\mxp[AC]=0$ since the only mathematical expectation which is not null, $\mxp[A^2]$, is multiplied by $\mxp[CD]=0$
    	\item $\mxp[A^3B]=3\mxp[A^2]\cdot\mxp[BC]=0$ since $\mxp[BC]=0$\vspace*{2pt}
    	\item $\begin{aligned}
                \mxp[A^2B^2]&=\mxp[A^2]\cdot\mxp[B^2]+2\mxp^2[AB]\\
                &=\mxp[A^2]\cdot\mxp[B^2]\neq 0.
              \end{aligned}$
        \item $\begin{aligned}
                \cov[A^2B^2] &=\mxp[A^2B^2]-\mxp[A^2]\cdot\mxp[B^2]\\
                &=\mxp[A^2]\cdot\mxp[B^2]-\mxp[A^2]\cdot\mxp[B^2]=0.
              \end{aligned}$
    \end{itemize}

	   \subsubsection{Variance of $\widehat{S_{\mathrm{cs}}}$}
    From (\ref{eq:def_cs}), it comes
    $$
    \begin{aligned}
        \widehat{S_{\mathrm{cs}}} = \displaystyle \frac{1}{\binom{q}{2}} &\left[ \sum_{j=1}^{q-1}\sum_{k=j+1}^{q} (N_j N_k + N_j' N_k') {\times}\right.\\
        & \quad {\times} (q-1) \sum_{j=1}^q (N_j S + N_j' R') {\times}\\
        &\qquad {\times} \left.\binom{n}{2} (R^2 + R'^2) \right].
    \end{aligned}
    $$
    Then,
    $$
    \begin{aligned}
        \var[\widehat{S_{\mathrm{cs}}}] = \displaystyle \frac{1}{\binom{q}{2}^2} &\left[ \sum_{j=1}^{q-1}\sum_{k=j+1}^{q} (\var[N_j N_k] + \var[N_j' N_k']) {\times}\right.\\
        & \quad {\times} (q-1)^2 \sum_{j=1}^q (\var[N_j R] + \var[N_j' R']) {\times}\\
        &\qquad {\times} \left.\binom{q}{2}^2 (\var[R^2] + \var[R'^2]) \right].
    \end{aligned}
    $$
    where all covariance terms are null thanks to Isserlis'theorem.
    From the properties (\ref{eq:prop1}) and (\ref{eq:prop2}), it comes
    $$
    \begin{aligned}
        \var[\widehat{S_{\mathrm{cs}}}] = \displaystyle \frac{1}{\binom{q}{2}^2} &\left[ \sum_{j=1}^{q-1}\sum_{k=j+1}^{q} (\var[N_j] \var[N_k] + \var[N_j'] \var[N_k']) {\times}\right.\\
        & \quad {\times} (q-1)^2 \sum_{j=1}^q (\var[N_j] \var[R] + \var[N_j'] \var[R']) {\times}\\
        & \qquad {\times} \left.\binom{q}{2}^2 (2\var^2[R] + 2\var^2[R']) \right].
    \end{aligned}
    $$
    Therefore,
    \begin{equation}
        \begin{aligned}
            \var[\widehat{S_{\mathrm{cs}}}] &= \displaystyle \frac{1}{\binom{q}{2}^2} \left[ 2\binom{q}{2} \frac{\sigma_N^4}{4} + 2q(q-1)^2 \frac{\sigma_N^2\sigma_R^2}{4} + 4\binom{q}{2}^2\frac{\sigma_R^4}{4}\right]\\
            &= \frac{1}{q(q-1)} \sigma_N^4 + \frac{2}{q}\sigma_N^2\sigma_R^2 + \sigma_R^4.\label{eq:var_cs}
        \end{aligned}
    \end{equation}

    	\subsubsection{Variance of $\widehat{S_{\mathrm{sa}}}$}
    From (\ref{eq:def_sa}), it comes
    $$
    \begin{aligned}
        \widehat{S_{\mathrm{sa}}} = \frac{1}{q^2} &\left[ \sum_{j=1}^q (N_j^2+N_j'^2) + q^2 (R^2 + R'^2) +\right.\\
        & \quad + 2\sum_{j=1}^{q-1}\sum_{k=j+1}^{q} (N_j N_k + N_j' N_k') +\\
        & \qquad + \left. 2q\sum_{j=1}^{q} (N_j R + N_j' R')\right]
    \end{aligned}
    $$
    Then,
    $$
    \begin{aligned}
        \var[\widehat{S_{\mathrm{sa}}}] = \frac{1}{q^4} &\left[ \sum_{j=1}^q (\var[N_j^2]+\var[N_j'^2]) +  q^4 (\var[R^2] + \var[R'^2]) +\right.\\
        & \quad + 4\sum_{j=1}^{q-1}\sum_{k=j+1}^{q} (\var[N_j N_k] + \var[N_j' N_k']) + \\
        & \qquad + \left. 4q^2\sum_{j=1}^{q} (\var[N_j R] + \var[N_j' R'])\right]
    \end{aligned}
    $$
    where all covariance terms are null thanks to Isserlis'theorem.
    From the properties (\ref{eq:prop1}) and (\ref{eq:prop2}), it comes
    $$
    \begin{aligned}
        \var[\widehat{S_{\mathrm{sa}}}] = \frac{1}{q^4} &\left[ \sum_{j=1}^q (2\var^2[N_j]+2\var^2[N_j']) + q^4 (2\var^2[R] + 2\var^2[R']) +\right.\\
        & \quad + 4\sum_{j=1}^{q-1}\sum_{k=j+1}^{q} (\var[N_j] \var[N_k] + \var[N_j'] \var[N_k']) + \\
        & \qquad + \left. 4q^2\sum_{j=1}^{q} (\var[N_j] \var[R] + \var[N_j'] \var[R'])\right]
    \end{aligned}
    $$
    Therefore,
    \begin{equation}
        \begin{aligned}
            \var[\widehat{S_{\mathrm{sa}}}] &= \displaystyle \frac{1}{q^4} \left[ 4q\frac{\sigma_N^4}{4} + 4q^4\frac{\sigma_R^4}{4}+ 8\binom{q}{2}\frac{\sigma_N^4}{4} + 8q^3\frac{\sigma_N^2\sigma_R^2}{4} \right]\\
            &= \frac{1}{q^2} \sigma_N^4 + \frac{2}{q}\sigma_N^2\sigma_R^2 + \sigma_R^4.\label{eq:var_sa}
        \end{aligned}
    \end{equation}

    \subsubsection{Variance ratios}
    Let us compare the cross-spectrum and spectrum average estimates variances for limit signal to noise ratio values.\\
    If $\sigma_R^2\ll \sigma_N^2$,
    $$
    \var[\widehat{S_{\mathrm{cs}}}]\approx \frac{1}{q(q-1)} \sigma_N^4 \qquad \textrm{and} \qquad \var[\widehat{S_{\mathrm{sa}}}]\approx \frac{1}{q^2} \sigma_N^4.
    $$
    Consequently,
    $$
    \var[\widehat{S_{\mathrm{cs}}}]\approx \frac{q}{q-1} \var[\widehat{S_{\mathrm{sa}}}].
    $$

    If $\sigma_N^2\ll \sigma_R^2$},
    $$
    \var[\widehat{S_{\mathrm{cs}}}]\approx \sigma_R^4 \qquad \textrm{and} \qquad \var[\widehat{S_{\mathrm{sa}}}]\approx \sigma_R^4.
    $$
    Consequently,
    $$
    \var[\widehat{S_{\mathrm{cs}}}]\approx \var[\widehat{S_{\mathrm{sa}}}].
    $$

\end{document}